\documentclass[12pt]{article}
\newtheorem{lemma}{Lemma}
\newtheorem{proposition}{Proposition}

\newtheorem{definition}{Definition}
\def\tr{\mathop{\rm Tr}\nolimits}

\def\di{\mathop{\textrm{d}}\nolimits}

\def\dif{\mathop{\rm d}\nolimits}
\def\ci{\mathop{\textrm{i}}\nolimits}

\catcode`\@=11
\long\def\@makefntext#1{
\protect\noindent \hbox to 3.2pt {\hskip-.9pt
$^{{\ninerm\@thefnmark}}$\hfil}#1\hfill}        %CAN BE USED

 \def\@makefnmark{\hbox to 0pt{$^{\@thefnmark}$\hss}}  %ORIGINAL

\def\ps@myheadings{\let\@mkboth\@gobbletwo
\def\@oddhead{\hbox{}
\rightmark\hfil\ninerm\thepage}
\def\@oddfoot{}\def\@evenhead{\ninerm\thepage\hfil
\leftmark\hbox{}}\def\@evenfoot{}
\def\sectionmark##1{}\def\subsectionmark##1{}}

\renewenvironment{thebibliography}[1]
    {\begin{list}{$^{\arabic{enumi}}$}
    {\usecounter{enumi}\setlength{\parsep}{0pt}
\setlength{\leftmargin 1.25cm}{\rightmargin 0pt}
     \setlength{\itemsep}{0pt} \settowidth
    {\labelwidth}{#1.}\sloppy}}{\end{list}}

\newcounter{itemlistc}
\newcounter{romanlistc}
\newcounter{alphlistc}
\newcounter{arabiclistc}

\def\@citex[#1]#2{\if@filesw\immediate\write\@auxout
    {\string\citation{#2}}\fi
\def\@citea{}\@cite{\@for\@citeb:=#2\do
    {\@citea\def\@citea{,}\@ifundefined
    {b@\@citeb}{{\bf ?}\@warning
    {Citation `\@citeb' on page \thepage \space undefined}}
    {\csname b@\@citeb\endcsname}}}{#1}}

\newif\if@cghi
\def\cite{\@cghitrue\@ifnextchar [{\@tempswatrue
    \@citex}{\@tempswafalse\@citex[]}}
\def\citelow{\@cghifalse\@ifnextchar [{\@tempswatrue
    \@citex}{\@tempswafalse\@citex[]}}
\def\@cite#1#2{{$\null^{#1}$\if@tempswa\typeout
    {IJCGA warning: optional citation argument
    ignored: `#2'} \fi}}

\def\fnt#1#2{\footnotetext{\kern-.3em
    {$^{\mbox{\sevenrm #1}}$}{#2}}}

 1
 1
 1
\font\ninerm=cmr9

\begin{document}
\title{Type I vacuum solutions with aligned Papapetrou \\fields: an intrinsic characterization}
\vspace{1cm}
\author{Joan Josep Ferrando$^1$ and Juan Antonio S\'aez$^2$}
%\date{\today}
\date{\empty}
\vspace{1cm}
\maketitle
%\vspace*{-0.5cm}
\begin{abstract}
We show that Petrov type I vacuum solutions admitting a Killing
vector whose Papapetrou field is aligned with a principal bivector
of the Weyl tensor are the Kasner and Taub metrics, their
counterpart with timelike orbits and their associated windmill-like
solutions, as well as the Petrov homogeneous vacuum solution. We
recover all these metrics by using an integration method based on an
invariant classification which allows us to characterize every
solution. In this way we obtain an intrinsic and explicit algorithm
to identify them.

\begin{center}
PACS: 04.20.Cv, 04.20.-q
\end{center}

\end{abstract}

\vspace*{4cm}

\vspace*{2cm} \noindent $^1$ Departament d'Astronomia i
Astrof\'{\i}sica, Universitat de Val\`encia, E-46100 Burjassot,
Val\`encia, Spain.
E-mail: {\tt joan.ferrando@uv.es}\\
$^2$ Departament de Matem\`atiques per a l'Economia i l'Empresa,
Universitat de Val\`encia, E-46071 Val\`encia, Spain. E-mail: {\tt
juan.a.saez@uv.es}
\newpage

\section{Introduction}
If $\xi$ is a Killing vector, the {\it Killing 2--form} $\nabla \xi$
is closed and, in the vacuum case, it is a solution of the
source-free Maxwell equations. Because this fact was pointed out by
Papapetrou,\cite{pap} the covariant derivative $\nabla \xi$ has also
been called the {\it Papapetrou field}.\cite{faso1} In the Kerr
geometry the principal directions of the Killing 2--form associated
with the timelike Killing vector coincide with the two double
principal null (Debever) directions of the Weyl tensor.\cite{faso1}
This means that the Killing 2--form is a Weyl principal bivector.
This fact has been remarked upon by Mars\cite{mars} who has also
shown that it characterizes the Kerr solution under an asymptotic
flatness behavior.

A question naturally arises: can all the vacuum solutions with this
property of the Kerr metric be determined? In other words, is it
possible to integrate Einstein vacuum equations under the hypothesis
that the spacetime admits an isometry whose Killing 2--form is a
principal bivector of the Weyl tensor? Some partial results are
known about this question. Thus, we have studied the case of Petrov
type D spacetimes elsewhere\cite{fsDB} and we have shown that the
Kerr-NUT solutions are the type D vacuum metrics with a time-like
Killing 2--form aligned with the Weyl geometry.

Metrics admitting an isometry were studied by considering the
algebraic properties of the associated Killing
2--form,\cite{deb1,deb2} and this approach was extended to the
spacetimes with an homothetic motion.\cite{mc1,mc2} More recently
Fayos and Sopuerta\cite{faso1,faso2} have developed a formalism that
improves the use of the Killing 2--form and its underlined algebraic
structure for analyzing the vacuum solutions with an isometry. They
consider two new viewpoints that permit a more accurate
classification of these spacetimes: (i) the differential properties
of the principal directions of the Killing 2--form, and (ii) the
degree of alignment of the principal directions of the Killing
2--form with those of the Weyl tensor. The Fayos and Sopuerta
approach uses the Newman-Penrose formalism and several extensions
have been built for homothetic and conformal motions\cite{ste,lud}
and for non vacuum solutions.\cite{faso3}

Some of the conditions on the Killing 2--form imposed in the
literature quoted above could be very restrictive. Thus, in a
previous paper\cite{fsI1a} we have shown that the Petrov type I
vacuum spacetimes admitting an isometry whose Killing 2--form is
aligned with a Weyl principal bivector belong to two classes of
metrics which admit a 3--dimensional group of isometries of Bianchi
type I or II. In the present work we show that a close relation
between the Weyl principal directions and the isometry group exists
in these classes. This fact allows us to achieve an integration of
the vacuum equations with the aid of an invariant classification
and, in this way, to obtain an intrinsic and explicit
characterization of all the Petrov type I vacuum solutions that
admit an aligned Papapetrou field. It is worth remarking that the
integration method used here could be suitable in order to obtain
other type I solutions.

The vacuum homogeneous Petrov solution\cite{pet} was found to be the
only one satisfying: (i) vacuum, (ii) existence of a simply
transitive group $G_4$ of isometries. Although these two conditions
characterize the Petrov metric, it is quite difficult to know when a
metric tensor (given in an arbitrary coordinate system) satisfies
them. Indeed, the first condition is intrinsic because it imposes a
restriction on a metric concomitant, the Ricci tensor. Nevertheless,
the second one imposes equations that mix up, in principle, elements
other than the metric tensor (Killing vectors of the isometry group)
and, consequently, it can not be verified by simply substituting the
metric tensor. In Ref. 13 we have changed this last non intrinsic
condition for an intrinsic one: the Weyl tensor is Petrov type I
with constant eigenvalues. Moreover, as the Ricci and Weyl tensors
are concomitants of the metric tensor, $Ric \equiv Ric(g)$, ${\cal
W} \equiv {\cal W}(g)$, we have finally obtained the following {\it
intrinsic} and {\it explicit} characterization of the Petrov
solution:\cite{fsI1a} {\it the necessary and sufficient conditions
for $g$ to be the Petrov homogeneous vacuum solution are}
\begin{equation}
Ric = 0 \, ,  \qquad  6 (\tr{\cal W}^2)^3 \neq (\tr {\cal W}^3)^2
\, , \qquad \di \tr {\cal W}^2 = \di \tr {\cal W}^3 = 0
\end{equation}

A whole intrinsic and explicit characterization of a metric or a
family of metrics is quite interesting from a conceptual point of
view and from a practical one because it can be tested by direct
substitution of the metric tensor in arbitrary coordinates. Thus, it
is an approach to the metric equivalence problem alternative to the
usual one. This and other advantages have been pointed out
elsewhere\cite{fsS} where this kind of identification has been
obtained for the Schwarszchild spacetime as well as for all the
other type D static vacuum solutions. A similar study has been
fulfilled for a family of Einstein-Maxwell solutions that include
the Reissner-Nordstr\"{o}m metric.\cite{fsD}

In order to obtain intrinsic and explicit characterizations, as
well as having an intrinsic labelling of the metrics, we need  to
express these intrinsic conditions in terms of explicit
concomitants of the metric tensor. When doing this, the role
played by the results on the covariant determination of the
eigenvalues and eigenspaces of the Ricci tensor\cite{bcm} and the
principal 2--forms and principal directions of the Weyl
tensor\cite{fsI,fms} is essential.

In this work we solve vacuum equations under the hypothesis that the
spacetime is Petrov type I and there is a Killing vector whose
associated Papapetrou field is a eigenbivector of the Weyl tensor.
In this way, we recover the Petrov homogeneous vacuum solutions as
well as the Kasner and Taub metrics, their counterpart with timelike
orbits and their associated windmill-like solutions. Our integration
method is based on an invariant classification which allows us to
characterize the solutions intrinsically and explicitly. For every
solution some properties of the isometry group and the aligned
Killing 2--forms are given in terms of the Weyl principal
directions.

The article is organized as follows. In section 2 we present the
Cartan formalism adapted to the Weyl principal frame that a Petrov
type I space-time admits. In section 3 we summarize some results
needed here about type I vacuum metrics admitting aligned Papapetrou
fields. In section 4 we write vacuum Einstein equations for the
families of Petrov type I metrics that, having a non constant Weyl
eigenvalue, admit aligned Papapetrou fields. Sections 5 and 6 are
devoted to integrate these equations in different invariant
subcases, as well as to determine, for every solution, the Killing
vectors with an aligned Killing 2--form. In section 7 we present a
similar study when all the Weyl eigenvalues are constant. Finally,
in section 8, we summarize the results in an algorithmic form in
order to make the intrinsic and explicit character of our results
evident.

\section{Cartan formalism in the Weyl frame of a type I space-time}

The algebraic classification of the Weyl tensor was first tackled by
Petrov\cite{petrov-W} considering the number of the invariant
subespaces of the Weyl tensor regarded as an endomorphism of the
2-forms space. This classification was completed by
G\'eh\'eniau\cite{geh} and Bel\cite{be1} considering also the
eigenvalue multiplicity. In this framework appears the notion of
{\em Weyl principal bivector} that we use here and which was widely
analyzed by Bel\cite{be2} for the different algebraic types. In the
sixties many other authors presented alternative approaches to this
classification, and in more recent studies\cite{kra,fms} a wide
bibliography on this subject can be found. For short, we refer the
different classes of the Weyl tensor as the Petrov types. An
algebraically general Weyl tensor is Petrov type I.

In a Petrov type I spacetime the Weyl tensor $W$ determines four
orthogonal principal directions which define the {\em Weyl principal
frame} $\{ e_{\alpha} \}$.\cite{be2,fms} Then, the bivectors
(self-dual 2--forms) ${\cal U}_i =\frac{1}{2} \ (U_i - \ci *U_i ) $,
with $U_i = e_0 \wedge e_i $, are eigenbivectors of the self-dual
Weyl tensor ${\cal W} =\frac{1}{2} ( W - \ci * W)$, $*$ being the
Hodge dual operator. These bivectors satisfy $2 {\cal U}_i \times
{\cal U}_i =g$, where $\times$ denotes the contraction of adjacent
index in the tensorial product. The tern $\{ {\cal U}_i \}$
constitutes an orthonormal frame in the bivector space which has the
induced orientation given by
\begin{equation} \label{orient}
 {\cal U}_i \times {\cal U}_j = - \frac{\ci}{ \sqrt{2}} \epsilon_{ijk} \ {\cal U}_k  \, ,
\qquad \quad i \neq j
\end{equation}
If $\alpha_i$ is the eigenvalue associated with the eigenbivector
${\cal U}_i$, the self-dual Weyl tensor takes the canonical form
\begin{equation} \label{can1}
{\cal W} = - \sum_{i=1}^3 \alpha_i \, {\cal U}_i \otimes {\cal U}_i
\end{equation}
The Cartan formalism can be referred to the Weyl principal frame $\{
e_{\alpha} \}$ or, equivalently, to the {\em frame of
eigenbivectors} $\{ {\cal U}_i \}$. So, the six connection 1-forms
$\omega^{\beta}_{\alpha}$ defined by $\nabla e_{\alpha} =
\omega_{\alpha}^{\beta} \otimes e_{\beta}$  can be collected into
three complex ones $\Gamma_{i}^{j}$ ($\Gamma_i^j =-\Gamma_j^i$), and
the first structure equations take the expression:
\begin{equation}\label{es1a}
\nabla {\cal U}_i = \Gamma_i^j \otimes {\cal U}_j \, , \qquad
\Gamma_{i}^{j}= \omega_{i}^{j}   - \epsilon_{ijk} \ \omega_{0}^{k}
\, .
\end{equation}
The second  structure equations for a vacuum type I spacetime follow
by applying the Ricci identities $\nabla_{[\alpha} \nabla_{\beta]}
{{\cal U}_{i}}_{\, \epsilon \delta} = {{{\cal U}_{i}}_{\,
\epsilon}}^{\mu} R_{\mu \delta \beta \alpha} + {{\cal U}_{i}^{\
\mu}}_{\delta}  R_{\mu \epsilon  \beta \alpha} \, $, and in terms of
the eigenbivectors $\{ {\cal U}_i \}$ they  can be written as
\begin{equation} \label{es2a}
\di\Gamma_i^k -\Gamma_i^j \wedge \Gamma_j^k  = \ci \sqrt{2}
\epsilon_{ikm} \alpha_m {\cal U}_m
\end{equation}
If we make the product of each of these second structure equations
(\ref{es2a}) with ${\cal U}_m$ we can obtain the following three
complex scalar equations:
\begin{equation}\label{traza}
\nabla \cdot  \lambda_i = \lambda_{i}^2  -  (\lambda_{j}-
\lambda_{k})^2 -  \alpha_i \qquad (i, j , k \ \neq )
\end{equation}
where $\lambda_i = - {\cal U}_i (\nabla \cdot {\cal U}_i )$, and we
have denoted $\nabla \cdot \equiv \tr \nabla $ and $\lambda_i^2 =
g(\lambda_i , \lambda_i)$. The three complex 1-forms $\lambda_i$
contain the 24 independent connection coefficients as the
$\Gamma_i^j$ do. In fact, by using (\ref{orient}) and the first
structure equations (\ref{es1a}), both sets $\{ \Gamma_i^j \}$ and
$\{ \lambda_i \}$ can be related by
\begin{equation} \label{05}
\lambda_i \equiv - {\cal U}_i (\nabla \cdot {\cal U}_i ) =
-\frac{\ci }{\sqrt{2}}\ \epsilon_{ijk} \ {\cal U}_k (\Gamma_i^j )
\end{equation} And the inverse of these expressions say that for
different $i, j , k$
\begin{equation}\label{cap3123}
 {\cal U}_k(\Gamma_i^j )  = \frac{\ci}{\sqrt{2}} \, \epsilon_{ijk} \,
 (  \lambda_i +\lambda_j-\lambda_k ) \, , \qquad (i, j , k \ \neq )
\end{equation}
The Bianchi identities in the vacuum case state that the Weyl tensor
is divergence-free $\nabla \cdot {\cal W}=0$, and from (\ref{can1})
they can be written as
\begin{equation} \label{bianchia}
\di \alpha_i = (\alpha_j - \alpha_k ) ( \lambda_j - \lambda_k ) - 3
\alpha_i \lambda_i   \qquad (i, j , k \ \neq )
\end{equation}

Equations (\ref{bianchia}) show the relation that exists between the
gradient of the Weyl eigenvalues and the 1-forms $\lambda_i$ in the
vacuum case. This fact has suggested a classification of Petrov type
I spacetimes taking into account the dimension of the space that $\{
\lambda_i \}$ generate. More precisely,\cite{fsI1a}
\begin{definition}
We say that a Petrov type I spacetime is of class I$_a$ ($a=1, 2,
3$) if the dimension of the space that $\{ \lambda_i \}$ generate is
$a$.
\end{definition}

Differential conditions of this kind were imposed by Edgar
\cite{edgar2} on the type I spacetimes, and he showed that in the
vacuum case his classification also has consequences on the
functional dependence of the Weyl eigenvalues. We have slightly
modified the Edgar approach in order to obtain a classification that
is symmetric in the principal structures of the Weyl tensor. We
remark the invariant nature of this classification: it is based on
the vector Weyl invariants $\lambda_i$.

We have been studied elsewhere the symmetries of the vacuum metrics
of class $I_1$ and we have shown:\cite{fsI1a}
\begin{lemma} \label{lemma:G3}
A vacuum metric of class I$_1$ admits at least a (simply transitive)
group G$_3$ of isometries. It admits a G$_4$ if, and only if, it has
constant eigenvalues.
\end{lemma}

\section{Aligned Killing 2--forms and type I vacuum metrics}

If $\xi$ is a (real) Killing vector its covariant derivative $\nabla
\xi$ is named {\it Killing 2--form} or {\it Papapetrou
field}.\cite{pap,faso1} The Papapetrou fields have been used to
study and classify spacetimes admitting an isometry or an homothetic
or conformal motion (see Ref. 2-12). In this way, some classes of
vacuum solutions with a principal direction of the Papapetrou field
aligned with a (Debever) null principal direction of the Weyl tensor
have been considered.\cite{faso2} Also, the alignment between the
Weyl principal plane and the Papapetrou field associated with the
time-like Killing vector has been shown in the Kerr
geometry.\cite{mars,faso2}

Is it possible to determine all the vacuum solutions having this
property of the Kerr metric? Elsewhere\cite{fsDB} we give an
affirmative answer to this question for the case of Petrov type D
spacetimes by showing that {\it the type D vacuum solutions with a
time-like Killing 2--form aligned with the Weyl geometry are the
Kerr-NUT metrics}. In this work we accomplish this study for the
Petrov type I space-times by obtaining all the vacuum solutions with
this property and by determining the Killing vectors with an aligned
Killing 2--form. Moreover, we show the close relation between the
Weyl tensor geometry and the geometries of $\xi$ and $\nabla \xi$.

In order to clarify what kind of alignment between the Killing
2--form and the Weyl tensor is analyzed in this work we give the
specific definition.  If $\{ {\cal U}_i \}$ is an orthonormal basis
of the self-dual 2-forms space, the Papapetrou field $\nabla \xi$
associated to a Killing vector $\xi$ has, generically, three
independent complex components $\Omega_i$:
\begin{equation} \label{ka1}
\nabla \xi = \sum_{i=1}^3 \Omega_i {\cal U}_i + \sum_{i=1}^3
\tilde{\Omega}_i \tilde{\cal{U}}_i
\end{equation}
where $\tilde{}$ means complex conjugate. Then:
\begin{definition}
We say that a Papapetrou field $\nabla \xi$ is aligned with a
bivector ${\cal U}$ if both 2--forms have the same principal
2--planes, that is, $\nabla \xi = \Omega {\cal U} + \tilde{\Omega}
\tilde{\cal U}$. \\
We say that a Papapetrou field $\nabla \xi$ is aligned (with the
Weyl tensor) if it is aligned with a Weyl principal bivector.
\end{definition}

When a Killing 2--form is aligned with a bivector of an orthonormal
frame of invariants bivectors ${\cal U}_i$, the Killing vector is
strongly restricted by the connection 1--forms. Thus, for type I
metrics we have:\cite{fsI1a}
\begin{lemma} \label{lemma-conexio}
In a Petrov type I spacetime with a Killing vector $\xi$, the
Papapetrou field $\nabla \xi$ is aligned with a Weyl principal
2-form ${\cal U}_i $ if, and only if, $\xi$ is orthogonal to the two
complex connection 1-forms $\Gamma_i^j$ (defined by the Weyl
principal frame $\{ {\cal U}_i \}$).
\end{lemma}
The alignment between a Killing 2-form and a Weyl principal bivector
of a type I vacuum solution has been partially analyzed\cite{fsI1a}
and the following necessary condition has been obtained:
\begin{lemma}  \label{lemma-aligned-I1}
A vacuum Petrov type I spacetime which admits a Killing field with
an aligned Papapetrou field belongs to class I$_1$.
\end{lemma}

As a consequence of lemmas \ref{lemma:G3} and
\ref{lemma-aligned-I1}, we obtain that {\em a vacuum Petrov type I
spacetime which admits a Killing field with an aligned Papapetrou
field admits, at least, a 3-dimensional group of isometries.} This
means that a unique symmetry with an aligned Papapetrou field
implies that other symmetries exist.

These results imply that in order to find all the type I vacuum
solutions admitting an aligned Papapetrou field, we must analyze the
vacuum solutions of class I$_1$. We shall start with the case where
a non constant eigenvalue $\alpha_1$ exists. After that we shall
finish by dealing with the case of all the eigenvalues being
constant.

\section{Vacuum equations for the class I$_1$}

As we know that every vacuum solution of class I$_1$ admits a
(simply transitive) $G_3$ group of isometries, a real function
$\tau$ exists such that $\alpha_i \equiv \alpha_i (\tau)$.
Moreover, as we are in class I$_1$, it must be $\lambda_i \wedge
\lambda_j =0$ and so, from Bianchi identities (\ref{bianchia}), we
obtain $\lambda_i \wedge \di \alpha_1 =0$. So, taking into account
(\ref{05}) and that a G$_3$ is admitted, three functions
$\varphi_i (\tau)$ exist such that
 \begin{equation}\label{cero}
\Gamma_i^j = \ci \,  \epsilon_{ijk} \,\varphi_k \, {\cal U}_k ( \di \tau)
\end{equation}
On the other hand, it has also been shown in Ref. 13 that $\di
\alpha_1$ can not be a null vector and so, $ (\di \tau)^2 \neq 0$.
Thus, $\{ \di \tau , u^i \}$, with $u^i = {\cal U}_i (\di \tau )$,
is an orthogonal frame such that $2(u^i)^2 = -(\di \tau)^2$ . Then,
we can write the bivectors $\{ {\cal U}_i \}$ as
 \begin{equation}\label{ui}
 {\cal U}_i = - \frac{1}{{ (\di \tau )}^2}
 \left( {\di \tau} \wedge u^i \ + \frac{\ci}{ \sqrt{2}} \
\epsilon_{ijk} \, u^j \wedge  u^k \right)
\end{equation}
We can use this expression to eliminate ${\cal U}_i $ in the second
structure equations (\ref{es1a}) and then they become an exterior
system for the orthonormal frame $\{ \di \tau , u^i \}$:
\begin{equation}\label{es2def}
\di u^{i} = \mu_i (\tau) \ \di \tau \wedge u^{i} + \nu_i (\tau) \
u^j \wedge u^k
\end{equation}
for every cyclic permutation $i,j,k$, and where the functions
$\mu_i$ and $\nu_i$ are given by:
\begin{equation} \label{mus}
\mu_i = - (\ln \varphi_i)' + \frac{\sqrt{2}
\alpha_i}{\varphi_i \ (\di \tau)^2 } ; \ \ \ \ \
\nu_i = - \ci \Big( \frac{2 \alpha_i}{\varphi_i \ (\di \tau)^2} +
\frac{ \varphi_j \varphi_k }{\varphi_i}  \Big)
\end{equation}
where $^{\prime}$ stands for the derivative with respect to the
variable $\tau$. But $\di \tau$ is proportional to the invariant
1--form $\di \alpha_1$ and a G$_3$ exists. Thus, it follows that
$(\di \tau)^2$ and $\Delta \tau$ depend on $\tau$. This fact
allows us to choose $\tau$ such that $\Delta \tau =0$. Then,
$\tau$ is fixed up to an affine transformation $\tau
\hookrightarrow \alpha \tau + \beta $. In terms of this harmonic
function, the equations (\ref{traza}) become:
\begin{equation}\label{trazadef}
\Big( (\varphi_j + \varphi_k)^{\prime} - \sqrt{2}  \varphi_j
\varphi_k\Big) (\di \tau)^2 = 2 \sqrt{2} \alpha_i
\end{equation}
for every cyclic permutation of $i,j,k$.
The Bianchi identities  (\ref{bianchia}) can be
stated as
\begin{equation}\label{bianchi}
\hspace{-0.6cm} \alpha_1^{\prime} = \frac{1}{\sqrt{2}} \Big(
\varphi_3 ( \alpha_2 - \alpha_1 ) - \varphi_2 (2 \alpha_1 +
\alpha_2) \Big)  , \ \   \alpha_2^{\prime} = \frac{1}{\sqrt{2}}
\Big( \varphi_3 ( \alpha_1 - \alpha_2) - \varphi_1 (\alpha_1 + 2
\alpha_2) \Big)
\end{equation}

At this point, it is clear that the integration of the system
(\ref{es2def}) depends strongly on the number of the $u^i = {\cal
U}_i (\di \tau )$ that are integrable 1--forms. Thus, it seems
suitable to give a classification of type $I_1$ spacetimes that
takes into account these restrictions. But these conditions lead to
an invariant classification because of $u^i$ is proportional to
${\cal U}_i(\mbox{d} \alpha_1) $:
\begin{definition} \label{subclass}
We will say that a Petrov type I$_1$  vacuum metric with $\mbox{\rm
d} \alpha_1 \neq 0$ is of class I$_{1A}$ ($A=0,1,2,3$) if there are
exactly $A$ integrable 1--forms in the set $\{ {\cal U}_i (\mbox{\rm
d} \alpha_1) \}$.
\end{definition}

We have studied elsewhere\cite{fsI1a} the symmetries that the
different classes $I_{1A}$ admit, as well as necessary conditions
for the alignment of the associated Killing 2--forms with the Weyl
tensor. Here we will make use of the following result:
\begin{lemma}  \label{lemma-I-II}
If a vacuum Petrov type I spacetime admits a Killing field with an
aligned Papapetrou field then either it is the Petrov solution (that
has constant eigenvalues) or it is of class I$_{12}$ or I$_{13}$.
These classes admit an isometry group $G_3$ of Bianchi types II and
I, respectively.
\end{lemma}

Thus, in order to find the vacuum solutions with aligned Papapetrou
fields, we must consider the Petrov solution that admits a $G_4$ or
the classes I$_{12}$ and I$_{13}$ . Now we obtain the vacuum
solutions for these two classes with non constant eigenvalues. To
accomplish this goal, we will integrate the Bianchi identities
(\ref{bianchi}) and the scalar equations (\ref{trazadef}) taking
into account that in class I$_{13}$ all the functions $\nu_i$ given
in (\ref{mus}) are zero and two of them vanish in class I$_{12}$.
Finally, the second structure equations (\ref{es2def}) will be
integrated to obtain the 1-forms $u^{i}$ in terms of real
coordinates. After that, the metric tensor will be obtained as
\begin{equation}   \label{metrica1}
g = \frac{1}{(\di \tau)^2} \Big[ \di \tau \otimes \di \tau - 2
\ \sum_{i=1}^3 u^{i} \otimes u^{i} \Big]
\end{equation}

It is worth pointing out that in the spacetimes of type $I_1$ that
we are studying here there exist two outlined coframes, namely,
the Weyl principal coframe $\{\theta^{\alpha}\}$ and that defined
by $\{\di \tau, u^i\}$. We will see in following sections the
close relation between both coframes for the spacetimes in classes I$_{12}$
and I$_{13}$. This fact allows us to give intrinsic conditions
that label every Type I vacuum solution admitting an aligned
Papapetrou field.

\section{Vacuum solutions of Class I$_{13}$}

In class I$_{13}$ all of the 1-forms $u^{i}$ are integrable. Then
equations (\ref{es2def}) hold with $\nu_i =0$. Taking into account
(\ref{mus}), we can solve the equations (\ref{trazadef}) and
(\ref{bianchi}) to obtain
\begin{eqnarray}
\alpha_1 = b e^{2 a  \beta \tau} , \qquad \alpha_2 = k \alpha_1 ,
\qquad (\mbox{d}
\tau)^2 = \frac{b  e^{2 a \beta \tau}}{a^2 k (k+1)}  \label{una} \\
\varphi_3= \sqrt{2} k a  , \quad \varphi_2=- \sqrt{2}
a (k+1) , \quad \varphi_1= - \sqrt{2} k (k+1) a  \label{unabis}
\end{eqnarray}
where $a$,  $b$ and $\beta=1 + k + k^2$ are non zero constants and
$k$ is different from $1$, $-2$ and $-\frac{1}{2}$ because $g$ is
not of Petrov type D, and different from $-1$ and $0$ because none
of the Weyl eigenvalues vanishes as a consequence of the
Szekeres-Brans theorem.\cite{sze,brans1} The second structure
equations (\ref{es2def}) constitute an exterior system for the
1-forms $u^{i}\equiv {\cal U}_i (\di \tau)$. It implies that three
complex functions $\{ x^i \}$ exist such that
\begin{equation}\label{solucion}
\displaystyle
 u^1 =e^{{a} \tau} \, \di x^1 , \qquad
 u^2 =e^{{a k^2} \tau} \, \di x^2 ,
\qquad
 u^3  =e^{{a (1+k)^2 } \tau} \, \di
x^3
\end{equation}
From here and (\ref{metrica1}), we can obtain the metric tensor $g$
in complex coordinates.  In order to get real coordinates, another
fact is needed. As $\tau$ is a real function, it follows that $\di
\tau$, $(\di \tau)^2$ and $\nabla \di \tau$ must be also real. If we
compute  $\nabla \di \tau$ by using (\ref{metrica1}), (\ref{una})
and (\ref{unabis}) we obtain that, necessarily, either all of the
coefficients are real and $\di \tau$ coincides with one of the
principal directions $\theta^{\alpha}$, or two of the coefficients
are conjugate and $\di \tau$ takes the direction of one of the
bisectors $\theta^i \pm \theta^j$ of a spacelike principal plane. We
shall analyze every case, but we must take into account that as $\di
\tau \wedge \di \alpha_1 =0$, these conditions can also be written
in terms of the Weyl eigenvalue.

\subsection{$\di \alpha_1$ is a principal direction }

In this case $a$ and $k$ are real constants. We must remark that if
$\di \tau$ is a principal direction, then $u^{i}\equiv {\cal U}_i
(\di \tau)$ are so. But when $\di \tau$ coincides with the timelike
principal direction $\theta^0$, every $u^i$ is a real direction and,
if $\di \tau$ is a spacelike principal direction, some of them are
purely imaginary. Now we will analyze each case
in detail.\\[2mm]
{\it (i) Case $\di \alpha_1 \wedge \theta^0 =0$.} We have $\di
\tau = e_0 (\tau) \theta^0 $. Then $ u^{i} = \frac{1}{\sqrt{2}}
e_0 (\tau) \theta^i $, and so $\di x^i$ of (\ref{solucion}) are
real. If we take into account that the harmonic coordinate $\tau$
is defined up to affine transformation, the metric tensor
(\ref{metrica1}) in real coordinates takes the form of the Kasner
metric
\begin{equation}\label{kasner}
\hspace{-1cm} \displaystyle g = - e^{-2 \tau}  \di \tau^2 + e^{2
( \frac{1}{\beta} - 1) \tau}  (\di x^1)^2 + e^{2  (\frac{ k^2
}{\beta} -1) \tau} (\di x^2)^2 + e^{2 ( \frac{(1+k)^2}{\beta} -1)
\tau} (\di x^3)^2
\end{equation}
The coordinate transformation  $e^{- \tau} = t $ changes the
harmonic time to the proper time and gives us the usual expression
for this solution.\cite{kas,kra}

We must check whether there is a Killing field with an aligned
Papapetrou field. We have established\cite{fsI1a} that this
condition is equivalent to a Killing field to be orthogonal to two
of the complex connection 1-forms $\Gamma_i^j$ (see lemma
\ref{lemma-conexio}). The real Killing fields of this metric are
$\xi = k_1
\partial_{x^1} + k_2  \partial_{x^2}  + k_3
\partial_{x^3}$. As the connection 1-forms $\Gamma_i^j$ are
collinear with $u^{i}$ it follows that every Killing field
$\partial_i$ satisfies this condition, and so, we have three
Killing fields such that their Papapetrou fields are aligned with
the three principal 2-forms.\\[2mm]
{\it (ii) Case $ \di \alpha_1 \wedge \theta^1 =0$.}
Now, $\di \tau = e_1 (\tau) \ \theta^1 $, and so
 $\displaystyle (\di \tau)^2 > 0$.  In order to get real
coordinates we must take into account that in this case
$\sqrt{2}\, u^1 = -e_1(\tau) \theta^0$, $\sqrt{2}\, u^2 = \ci
e_1(\tau) \theta^3$ and $\sqrt{2}\, u^3 = \ci e_1(\tau) \theta^2$. And
so, the coordinates adapted to $u^2 $ and $u^3 $ are purely
imaginary $x^a = \ci y^a$ ($a=2,3$), $y^a$ being real functions.
Then, for the metric tensor $g$ we get a similar expression to the
one in the previous case, the only change being the causal
character of the gradient of the Weyl eigenvalue:
\begin{equation}\label{kasner2}
\hspace{-0.5cm} \displaystyle g =
 e^{-2 \tau}  \di \tau^2 -
e^{2  ( \frac{1}{\beta} - 1) \tau}  (\di x^1)^2 + e^{2  (\frac{ k^2
}{\beta} -1) \tau} (\di y^2)^2 + e^{2 ( \frac{(1+k)^2}{\beta} -1)
\tau} (\di y^3)^2
\end{equation}
This is the static Kasner metric.\cite{kra}

This solution admits three Killing fields $\partial_{x^1}$,
$\partial_{y^2}$ and $\partial_{y^3}$ such that their Papapetrou
fields are aligned with the three principal bivectors of the Weyl
tensor. This finishes the study of the cases in which the gradient
of the invariant $\alpha_1(\tau)$ is collinear with a principal
direction of the Weyl tensor. The following proposition summarizes
the main results.

\begin{proposition}
The Kasner metrics (\ref{kasner}), (\ref{kasner2}) are the only
Petrov type I$_{13}$ vacuum solutions where the gradient of the Weyl
eigenvalue is a principal direction of the Weyl tensor. \\[1mm]
The metrics of this family admit three Killing fields $\xi_i$
which are collinear with the three principal directions ${\cal
U}_i(\di \alpha_1)$, such that their Papapetrou fields $\nabla
\xi_i$ are aligned with the three principal bivectors of the Weyl
tensor ${\cal U}_i$.
\end{proposition}

\subsection{$\di \alpha_1 $ is not a principal direction}

As we have commented below, in this case $\di \tau$ must take the
direction of one of the bisectors of a spacelike principal plane,
say $\theta^2 + \theta^3$, $ \di \tau \propto \theta^2 + \theta^3$.
Then, $u^1 \propto \theta^3 - \theta^2$, $u^2 \propto \theta^0 + \ci
\theta^1$ and $u^3 \propto \theta^0 - \ci \theta^1$. Moreover,
$\nabla \di \tau$ is real if, and only if, $ a $ is real and $2 k= -
1 + \ci n $, $n$ being a non zero real constant because the metric is not of type D
and $n^2 \neq 3$ because
$\beta $ can not be zero.  Then, the
coordinate $x^1$ of (\ref{solucion}) must be purely imaginary, $x^1
= \ci x$, and $x^2$ and  $x^3$ must be conjugated functions, that is
$x^2 =y - \ci z $, $x^3 = y + \ci z$. Thus we get a real coordinate
system $\{ \tau, x, y , z \}$ and, from (\ref{metrica1}), we find
the following expression of the metric tensor:
\begin{equation}\label{kasner3}
\begin{array}{l}
\displaystyle g= \frac14 (3-n^2)^2 e^{\frac12 (3-n^2)\tau}
\di{\tau^2} +
e^{-\frac12 (1+n^2) \tau} \di x^2 + \\[3mm]
\displaystyle + e^{\tau} \left[ \cos (n \tau )  [\di
 z^2 - \di y^2 ]   - 2 \sin (n \tau) \di y \di z \right]
\end{array}
\end{equation}
This is the so called {\it windmill solution}.\cite{kra,windmill}

To see if an aligned Killing 2--form can exist in this spacetime, we
must look for a Killing field to be orthogonal to two of the
connection 1-forms. The Killing fields of this solution are $\xi=
k_1 \partial_x + k_2 \partial_y + k_3 \partial_z$ and, as every
connection 1-form is parallel to one of the directions $u^{i}$, the
only Killing field which is orthogonal to a pair of connection
1-forms is $\partial_x$, that can be characterized as the Killing
field that takes the direction of the bisector $\theta^2 -
\theta^3$.
Moreover, the Weyl tensor has just a real eigenvalue $\alpha_1$ and if
${\cal U}_1$ is the associated eigenbivector, then ${\cal U}_1 (\dif \alpha_1 )$
is collinear with the Killing field $\partial_x$.
We can
collect these results in the following

\begin{proposition}
The windmill solution (\ref{kasner3}) is the only Petrov type
I$_{13}$ vacuum solution where the gradient of the Weyl eigenvalue
$\alpha_1$ is not a principal direction of the
Weyl tensor. \\
In such spacetime a unique real eigenvalue $\alpha_1$ exists. Then,
if ${\cal U}_1$ is the associated eigenvibector, the field
${\cal U}_1 (\dif \alpha_1 )$ is collinear with a Killing field that has a Papapetrou field
aligned with ${\cal U}_1$.

\end{proposition}

\section{Vacuum solutions of class $I_{12}$}

Let us suppose now that only two directions, let us say $u^2$ and
$u^3$, are integrable. So, we can take  $\nu_2=\nu_3=0$ in the
second structure equations  (\ref{es2def}). Taking into account
the definition of $\nu_i$ from (\ref{mus}) we obtain
\begin{equation}  \label{121}
\varphi_3=\frac{k}{\sqrt{2}} - \varphi_1 , \quad \quad \varphi_2=
\frac{a^2}{\sqrt{2}k} -  \varphi_1 , \quad \quad \varphi_1=
\frac{a}{\sqrt{2}} \frac{ b e^{- a \tau} + 1}{b e^{- a \tau} - 1}
\end{equation}
where  $a$, $b$ and $k$ are complex constants, $a^2 \not= k^2$. Then, by also using
the Bianchi identities (\ref{bianchi}) we obtain
\begin{equation} \label{122}
\displaystyle (\di \tau)^2  = \frac{- 2 \sqrt{2} c}{a}
e^{-  \frac{a^2 + a k + k^2}{k} \tau}
(b^2 e^{- 2  a \tau} - 1)^{-1}
\end{equation}
where $c$ is another complex constant.

As in the previous section, the only possibilities for $\nabla \di \tau$
to be real are that either $\di \tau$ is a principal direction
$\theta^{\alpha}$ or it is the bisector $\theta^i + \theta^j$ of a
spacelike principal plane.

\subsection{$\di \alpha_1$ is a principal direction}

In this case we have that $k$, $a^2$ and
$\frac{\varphi_3^{\prime}}{\varphi_3} - \sqrt{2} \varphi_3 $ are
real. From (\ref{121}) we obtain
\begin{equation}\label{123}
\frac{\varphi_3^{\prime}}{\varphi_3} - \sqrt{2} \varphi_3  = -
 a \frac{ b^2 +e^{2
\sqrt{2}\tau }}{b^2 - e^{2 \sqrt{2}\tau}}
\end{equation}
So we can  conclude that $a$ and $b^2$ must be real constants. Now
we shall go on the integration of (\ref{es2def}). As in the
previous section it will be useful to distinguish the cases of
$\di \tau$ to be the timelike principal direction $\theta^0$ or a
spacelike principal direction $\theta^i$. We will study these
cases separately.\\[1mm]
{\it (i) Case $\di \alpha_1 \wedge \theta^0 =0$.} Here we have
$\di \tau= e_0 (\tau) \theta^0 $, and so $ u^i $ must be real for
every $i$. Consequently, if we take into account (\ref{es2def})
with $\nu_2 = \nu_3 = 0$, real coordinates  $\{ x, y, z \}$ can be
found such that
\begin{equation}\label{37}
\hspace{-1.5cm} u^2= \frac{e^{-\frac{k}{2}  \tau }}{\sqrt{2}} \  \ \di x \, , \quad
u^3 = \frac{e^{- \frac{a^2}{2 k} \tau}}{\sqrt{2}} \  \ \di y \, , \quad u^1 = -
\frac{\ci \ \sqrt{2} a b e^{-  a \tau}}{b^2 e^{-2  a \tau}
-1} \ e^{- \frac{a^2 + k^2 }{ 2k} \tau} \ (\di z + x \di y)
\end{equation}
As $u^1$ is real, we find that $b$ is purely imaginary, $b = - \ci
\beta$. Then, from (\ref{122}) we can calculate $(\di \tau)^2$, and
taking into account the freedom of an affine transformation in
choosing the harmonic coordinate $\tau$nwe can take $\beta =1$ and
we can write the metric
in the usual form of the Taub\cite{ta2} metric
\begin{equation} \label{taub}
\hspace{-1.2cm} g =  \frac{\mbox{cosh} (a \tau)}{ a }  \left(  -
e^{ \frac{a^2 + k^2}{k} \tau} \,  \di \tau^2 + e^{\frac{a^2}{k}
\tau} \, \di x^2 + e^{ k \tau} \, \di y^2   \right)  +  \frac{
a}{\mbox{cosh} ( a \tau)} (\di z + x \di y )^2
\end{equation}

To see if a Killing field with an aligned Papapetrou field exists,
we must look for a Killing field which is orthogonal to two of the
connection 1-forms. The Killing fields of the Taub metric
(\ref{taub}) are $\xi = k_1 \partial_x + k_2 \partial_y + (k_3 -
k_1 y) \partial_z$ and, taking into account that the connection
1-forms $\Gamma_i^j$ are collinear with $u^k$, from (\ref{37}) we
find that the only Killing field that is orthogonal to a pair of
connection 1-forms is $\xi = \partial_z$, and it is orthogonal to
$\Gamma_1^2$ and $\Gamma_1^3$. So, the principal 2-form aligned
with a Papapetrou field is ${\cal U}_1$, and it is characterized
by the fact that ${\cal U}_1 (\di \tau)$ is not integrable.\\[1mm]
{\it (ii) Case $\di \alpha_1 \wedge \theta^1 =0$.} Now, $\di \tau
= e_1 (\tau) \theta^1$ and we have  $\sqrt{2}\, u^1 = -e_1(\tau)
\theta^0$, $\sqrt{2}\, u^2 = \ci e_1(\tau) \theta^3$ and
$\sqrt{2}\, u^3 = \ci e_1(\tau) \theta^2$. So, we can consider
real coordinates $\{ x , y , z \} $ such that
\begin{equation}
\hspace{-1.5cm} u^2  = \frac{\ci \  e^{- \frac{k}{2} \tau}}{ \sqrt{2}} \ \di x \, ,
\quad  u^3 = \frac{\ci e^{- \frac{a^2}{2 k } \tau}}{\sqrt{2}} \ \di y \, , \quad  u^1
= \frac{-\sqrt{2}  \beta a e^{-a \tau}}{\beta^2 e^{- 2  a \tau} + 1} e^{-\frac{a^2 +
k^2}{2 k } \tau} \ (\di z - x \di y)
\end{equation}
Then, the same analysis of the previous case leads to the
counterpart with timelike orbits of the Taub metric:\cite{kra}

\begin{equation}\label{taub2}
\hspace{-1cm} g =  \frac{\mbox{cosh} ( a \tau)}{ a }  \left(
e^{\frac{a^2 + k^2}{k} \tau} \,  \di \tau^2 + e^{\frac{a^2}{k}
\tau} \, \di x^2 + e^{k \tau} \ \di y^2   \right) - \frac{
a}{\mbox{cosh} (a \tau)} (\di z - x \di y )^2
\end{equation}

The same property of the Taub metric  concerned with the aligned
Papapetrou fields holds in this case. We can summarize these
results for the case that $\di \alpha_1$ is collinear with a
principal direction in the following

\begin{proposition}
The Taub metric (\ref{taub}) that has spacelike orbits, and its
counterpart with timelike orbits (\ref{taub2}) are the only type
I$_{12}$ vacuum solutions where the gradient of the Weyl
eigenvalue $\alpha_1$ is collinear with a principal
direction of the Weyl tensor.\\[1mm]
Both metrics admit a principal 2-form ${\cal U}_i$ such that
${\cal U}_i (\di \alpha_1)$ is not integrable. Then, the Killing
field collinear with ${\cal U}_i (\di \alpha_1)$  is the only one
whose Papapetrou field is aligned (with the principal 2-form
${\cal U}_i$).
\end{proposition}

\subsection{$\di \alpha_1 $ is not a  principal direction}

In this case $\di \tau$ must take the
direction of one of the bisectors of a spacelike principal plane,
say $\theta^2 + \theta^3$, $ \di \tau \propto \theta^2 +
\theta^3$. Then,  a
similar analysis to the one in the previous cases, leads to the
metric
\begin{equation}\label{taub3}
\begin{array}{l}
g=e^{2  m \tau} \left[ \frac{\mbox{cosh} (a \tau)}{a}    \di \tau^2 +
\frac{a}{\mbox{cosh} ( a \tau)} e^{-2  m \tau} (\di z - u \di v)^2 + \right. \\[3mm]
\left. \qquad + \frac{\mbox{cosh} ( a \tau)}{a} e^{-  m \tau}
\left[ \  \mbox{cos }(n   \tau)
(\di v^2 - \di u^2 ) - 2 \sin ( n \tau) \di u \di v
\right] \right]
\end{array}
\end{equation}
where  $a^2 =m^2 + n^2$, $n \not=0$. This is an equivalent {\it windmill-like}
metric for the Taub solution.

The real Killing  fields of this metric in the previous coordinate
system are $\xi=k_1 \partial_u + k_2 \partial_v + (k_1 v + k_3)
\partial_z$. As the complex connection 1-forms $\Gamma_i^j$ are
collinear with $u^k \equiv {\cal U}_k (\di \tau)$, we conclude
that there is only one Killing field $\partial_z $ which is
orthogonal to two connection 1--forms, more precisely, to
$\Gamma_1^2 $ and $\Gamma_1^3$. So, this Killing field has a
Papapetrou field which is aligned with the principal bivector
${\cal U} _1$. We summarize these results in the following
\begin{proposition}
The metric (\ref{taub3}) is the only vacuum solution of class
I$_{12}$ where the gradient of the Weyl eigenvalue
$\alpha_1$ is not a Weyl principal direction. \\[1mm]
This solution admits a principal 2-form ${\cal U}_i$ such that
${\cal U}_i (\di \alpha_1)$ is not integrable. Then, the Killing
field collinear with ${\cal U}_i (\di \alpha_1)$ is the only one
whose Papapetrou field is aligned (with the principal 2-form
${\cal U}_i$).

\end{proposition}

\section{Type I vacuum solutions with constant eigenvalues}

Elsewhere\cite{fsI1a} we have shown that the only Petrov type I
vacuum solution with constant eigenvalues is the homogeneous Petrov
solution.\cite{pet,kra} In real coordinates this metric writes as
\begin{equation}\label{pes}
\hspace{-1cm}
k^2 \ g = \di x^2 + e^{-2x} \di y^2 + e^x
\left( {\cos }\sqrt{3} x \left( \di z^2 - \di t^2 \right) - 2
\mbox{sin}\sqrt{3} x \di z \di t \right)
\end{equation}
The eigenvalues of this metric are proportional to the three cubic
roots of $-1$, $\alpha_i = k^2 \ \sqrt[3]{-1}$. So, a real
eigenvalue, let us say $\alpha_3$, exists. From the metric
expression (\ref{pes}) we get that $\Gamma_1^3 \wedge {\cal U}_1
(\Gamma_1^2)=0$ and $\Gamma_2^3 \wedge {\cal U}_2 (\Gamma_1^2)=0$.
Moreover, a straightforward calculation shows that $\di x$ takes the
direction of one of the bisectors of the plane $*U_3$, $\di x =
e_1(x) (\theta^1 + \theta^2)$, and that the complex connection
1-forms $\Gamma_i^j$ are give by
$$
\begin{array}{ll}
\Gamma_1^2 = e^{-x} \, \di y \, , \quad &
{\cal U}_3 (\Gamma_1^2) = - \frac{\ci}{\sqrt{2}} \, \di x \, , \\[3mm]
{\cal U}_1 ( \Gamma_1^2) = \frac{1}{2} e^{\frac{1}{2} (1 + \ci
\sqrt{3} ) x } \, (\di t - \ci \di z ) \, , \qquad & {\cal U}_2 (
\Gamma_1^2) = \frac{1}{2} e^{\frac{1}{2} \, (1 - \ci \sqrt{3} ) x
} \, (\di t + \ci \di z )
\end{array}
$$
The Killing fields of this solution are $\{\partial_t , \, \partial_z, \,
\partial_y ,\,\partial_x + y \partial_y + \frac{1}{2} ( \sqrt{3} t - z )
\partial_z - \frac{1}{2} (t + \sqrt{3} z ) \partial_t \}$ and so, it easily
follows:
\begin{proposition}
The Petrov homogeneous vacuum solution (\ref{pes}) admits just a
Killing field such that its Papapetrou field is aligned with a
principal bivector.
If $\alpha_3$ is the real eigenvalue, this Killing field is
proportional to $\Gamma_1^2$ and its Papapetrou field is aligned with ${\cal U}_3$.
\end{proposition}

\section{Summary in algorithmic form}

In this paper we have found all the Petrov type I vacuum solutions
admitting a Killing field whose Papapetrou field is aligned with a
principal bivector of the Weyl tensor. We knew\cite{fsI1a} that
these solutions admit either a simply transitive group $G_4$ of
isometries and then the metric must be  the homogeneous Petrov
solution (\ref{pes}), or a simply transitive $G_3$ group of
isometries and then the spacetime belongs to one of the classes
$I_{13}$ and $I_{12}$ in definition \ref{subclass}. Here we have
shown that these necessary conditions given in Ref. 13 are  also
sufficient conditions.

The solutions can be characterized by a condition on the normal
direction to the orbits group: for class $I_{13}$, (i) if it is a
timelike principal direction we reach the Kasner metric
(\ref{kasner}), (ii) if it is a spacelike principal direction we
reach the static Kasner metric (\ref{kasner2}), and (iii) if it is
not a principal direction we obtain the windmill Kasner metric
(\ref{kasner3}); for class $I_{12}$, under similar conditions, we
obtain (i) the Taub metric (\ref{taub}), (ii) the timelike
counterpart of the Taub metric (\ref{taub2}), and (iii) the
windmill-like metric for the Taub solution (\ref{taub3}).

It is worth pointing out that the integration procedure is based on
intrinsic conditions imposed on algebraic and differential
concomitants of the Weyl tensor. On the other hand, these Weyl
invariants can be obtained directly from the components of the
metric tensor $g$ in arbitrary local coordinates and without solving
any equations.\cite{fsI,fms} Consequently, we get an intrinsic and
explicit labelling of every solution (similar to that given for the
Petrov metric in Ref. 13). The table 1 summarizes these results and
enables us to obtain the directions of the Killing fields having
aligned Papapetrou field. In the table we find the Weyl tensor
invariants
\begin{eqnarray}
\hspace{-0.5cm} \alpha_i \equiv \alpha_i (g) \, , \qquad
\theta^{\alpha} \equiv \theta^{\alpha} (g) \, , \qquad {\cal U}_i
\equiv {\cal U}_i (g) \, , \label{con} \\ [1mm] \hspace{-0.5cm}
\lambda_i \equiv \lambda_i(g) = -{\cal U}_i (\nabla
\cdot {\cal U}_i) \, ,\label{u34} \\[1mm]
\hspace{-0.5cm} N \equiv N(g) \, , \  \mbox{\rm number of integrable
directions in the set} \  \{ {\cal U}_j (\di \alpha_i ) \}
\label{u35}
\end{eqnarray}
The metric concomitants (\ref{con}) are, respectively, the Weyl
eigenvalues $\alpha_i (g)$, the Weyl principal coframe
$\theta^{\alpha} (g)$ and the unitary Weyl principal bivectors
${\cal U}_i (g)$. The explicit expressions of these Weyl
invariants in terms of the Weyl tensor can be found
elsewhere.\cite{fsI,fms}

Finally,  to underline the intrinsic nature
 of our results we present a flow diagram that characterizes, among all the
vacuum solutions, those ones of type I having an aligned Papapetrou
field. This operational algorithm can be useful from a computational
point of view and  also involves the Weyl invariants (\ref{con}),
(\ref{u34}) and (\ref{u35}).

\begin{table}
\label{table1}

{\small
$$\begin{array}{|l|c|c|}
 \hline
  & \mbox{Intrinsic characterization}   & \mbox{Killing vectors with  }  \\[-2mm]
\mbox{SOLUTION}     &    &    \\[-4mm]
        &  \lambda_i \wedge \lambda_j =0 , \ \  \mbox{Ric } =0  & \mbox{aligned Papapetrou field  }    \\ \hline
         & \di \alpha_1 \neq 0  , \ \ N = 3  &       \\[-2mm]
 Kasner \  (21)      &             &      \\[-2mm]
         & \di \alpha_1 \wedge \theta^0 =0     &   \xi_i  \propto {\cal U}_i (\di \alpha_1), \  \ \ i= 1,2,3  \\ \cline{1-2}
%[-1mm]

    & \di \alpha_1 \neq 0  , \ \ N = 3  &   \nabla \xi_i \ \mbox{\ aligned with } \ {\cal U}_i     \\[-1mm]
 Kasner  \ (22)      &             &   \\[-2mm]
         & \di \alpha_1 \wedge \theta^j  =0 \ \  \mbox{for some } j    &      \\ \hline

   & \di \alpha_1 \neq 0  , \ \ N = 3  &  \exists{\rm !}  \   \alpha_{i_0} \; {\rm real },\ \ \xi \propto {\cal U}_{i_0} (\di \alpha_1)    \\[-2mm]

Windmill \ (23)      &             &    \\[-2mm]
         & \di \alpha_1 \wedge \theta^{\alpha } \neq 0 \ \  \forall \alpha   &  \nabla \xi \ \mbox{aligned with } {\cal U}_{i_0}  \\ \hline

         & \di \alpha_1 \neq 0  , \ \ N = 2 &       \\[-2mm]
 Taub \ (28)      &             &   \\[-2mm]
         & \di \alpha_1 \wedge \theta^0 =0     &   \\ \cline{1-2}
%[-1mm]

         & \di \alpha_1 \neq 0  , \ \ N = 2 &     \exists{\rm !}\  i_0 \ / \ {\cal U}_{i_0}(\di\alpha_1) \mbox{ is  not integrable}     \\[-2mm]
 Taub \ (30)      &             & \xi \ \propto \ {\cal U}_{i_0}(\di\alpha_1) \\[-2mm]
         & \di \alpha_1 \wedge \theta^j =0   \ \ \mbox{for some } j   &     \nabla \xi \mbox{ aligned with } {\cal U}_{i_0}     \\ \cline{1-2}

  & \di \alpha_1 \neq 0  , \ \ N =2  &     \\[-2mm]
Windmill \ (31)      &             &  \\[-2mm]
         & \di \alpha_1 \wedge \theta^{\alpha } \neq 0 \ \  \forall \alpha   &      \\ \hline

&    &       \exists{\rm !} \ \alpha_3  \  \mbox{real,} \   \ \xi \propto \Gamma_1^2\\[-2mm]
Petrov \ (32)      &     \di \alpha_i =  0  \ \ \forall i          &    \\[-2mm]
         &   &   \nabla \xi \mbox{ aligned with  } {\cal U}_3    \\ \hline

\end{array}$$

}

\caption{Type I vacuum solutions with aligned Papapetrou fields.}
\end{table}

\setlength{\unitlength}{0.9cm}
{\footnotesize \noindent
\begin{picture}(0,18)
\thicklines
\put(2,17){\vector(0,-1){1}}
\put(0,15){\line(2,1){2}}
\put(0,15){\line(2,-1){2}}
\put(4,15){\line(-2,1){2}}
\put(4,15){\line(-2,-1){2}}
\put(1.2,14.9){$\lambda_i \wedge \lambda_j =0$}
\put(4,15){\vector(1,0){1}}
\put(4,15){\vector(1,0){1}}

\put(4.4,15.1){no}
\put(2,14){\vector(0,-1){1}}
\put(2.1,13.4){yes}

\put(0,12){\line(2,1){2}}
\put(0,12){\line(2,-1){2}}
\put(4,12){\line(-2,1){2}}
\put(4,12){\line(-2,-1){2}}
\put(0.8,12){$\exists \, \alpha_i \ / \ \di \alpha_i \neq 0$}

\put(4,12){\vector(1,0){1}}
\put(2,11){\vector(0,-1){1}}
\put(4.4,12.1){no}
\put(2.1,10.5){yes}

\put(5,11.5){\line(0,1){1}}
\put(5,12.5){\line(1,0){3}}
\put(5,11.5){\line(1,0){3}}
\put(8,12.5){\line(0,-1){1}}
\put(5.3,11.8){Petrov (\ref{pes})}

\put(2,11){\vector(0,-1){1}}
\put(2,10){\line(-1,-1){1}}
\put(2,10){\line(1,-1){1}}
\put(2,8){\line(1,1){1}}
\put(2,8){\line(-1,1){1}}
\put(1.5,9){$N=3$}

\put(2,8){\vector(0,-1){1}}
\put(2.1,7.5){yes}

\put(2,7){\line(-2,-1){2}}
\put(2,7){\line(2,-1){2}}
\put(2,5){\line(-2,1){2}}
\put(2,5){\line(2,1){2}}
\put(0.9,6.1){$\di \alpha_i \wedge \theta^{\alpha} =0$ }
\put(0.8,5.7){(\footnotesize{for some} $\alpha$)}

\put(4,6){\vector(1,0){0.5}}
\put(4,6.1){no}
\put(2.1,4.7){yes}

\put(4.5,6.5){\line(0,-1){1}}
\put(4.5,6.5){\line(1,0){2.5}}
\put(4.5,5.5){\line(1,0){2.5}}
\put(7,5.5){\line(0,1){1}}
\put(4.5,5.9){{\it Windmill} (\ref{kasner3})}

\put(2,5){\line(0,-1){1}}
\put(2,4){\line(-2,-1){2}}
\put(2,4){\line(2,-1){2}}
\put(2,2){\line(-2,1){2}}
\put(2,2){\line(2,1){2}}
\put(0.9,2.9){$\di \alpha_i \wedge \theta^0 =0$}

\put(4,3){\vector(1,0){.5}}
\put(4,3.1){no}
\put(4.5,3.5){\line(1,0){2.5}}
\put(7,3.5){\line(0,-1){1}}
\put(4.5,2.5){\line(0,1){1}}
\put(4.5,2.5){\line(1,0){2.5}}
\put(4.6,2.9){Kasner (\ref{kasner2})}

\put(2,2){\vector(0,-1){1}}
\put(2.1,1.7){yes}
\put(2,1){\vector(1,0){2.5}}
\put(4.5,1.5){\line(0,-1){1}}
\put(4.5,1.5){\line(1,0){2.5}}
\put(7,0.5){\line(-1,0){2.5}}
\put(7,0.5){\line(0,1){1}}
\put(4.6,0.9){Kasner (\ref{kasner})}

\put(3,9){\vector(1,0){8}}
\put(3.8, 9.1){no}

\put(11,9){\line(1,1){1}}
\put(11,9){\line(1,-1){1}}
\put(13,9){\line(-1,1){1}}
\put(13,9){\line(-1,-1){1}}
\put(11.5, 8.9){$N=2$}
\put(12,8){\vector(0,-1){1}}

\put(12,8){\vector(0,-1){1}}
\put(12,7){\line(-2,-1){2}}
\put(12,7){\line(2,-1){2}}
\put(12,5){\line(-2,1){2}}
\put(12,5){\line(2,1){2}}

\put(10.9,6.1){$\di \alpha_i \wedge \theta^{\alpha} =0$ }
\put(10.8,5.7){(\footnotesize{for some} $\alpha$)}

\put(10,6){\vector(-1,0){0.6}}
\put(9.6,6.1){no}

\put(7.2, 6.5){\line(1,0){2.2}}
\put(7.2, 6.5){\line(0,-1){1}}
\put(9.4, 5.5){\line(-1,0){2.2}}
\put(9.4, 5.5){\line(0,1){1}}

\put(12,5){\vector(0,-1){1}}
\put(12,4){\line(-2,-1){2}}
\put(12,4){\line(2,-1){2}}
\put(12,2){\line(-2,1){2}}
\put(12,2){\line(2,1){2}}
\put(10.9,2.9){$\di \alpha_i \wedge \theta^0 =0$}

\put(10,3){\vector(-1,0){0.6}}
\put(7.2, 3.5){\line(1,0){2.2}}
\put(7.2, 3.5){\line(0,-1){1}}
\put(9.4, 2.5){\line(-1,0){2.2}}
\put(9.4, 2.5){\line(0,1){1}}

\put(12,2){\vector(0,-1){1}}
\put(12,1){\vector(-1,0){2.6}}

\put(7.2, 1.5){\line(1,0){2.2}}
\put(7.2, 1.5){\line(0,-1){1}}
\put(9.4, 0.5){\line(-1,0){2.2}}
\put(9.4, 0.5){\line(0,1){1}}

\put(7.4,5.9){Sol. (\ref{taub3})}
\put(7.4,2.9){Sol. (\ref{taub2})}
\put(7.4,0.9){Taub (\ref{taub})}
\put(11.35,7.5){yes}
\put(11.35,4.5){yes}
\put(11.35,1.5){yes}
\put(9.6,3.1){no}

\put(13,9){\vector(1,0){1}}
\put(13.3,9.1){no}

\put(0,17){\line(1,0){4}} \put(0,17){\line(0,1){1}}
\put(4,18){\line(-1,0){4}} \put(4,18){\line(0,-1){1}}
\put(1,17.4){$\alpha_i , \ \theta^{\alpha}, \ \lambda_i , \ N$}

\end{picture}
}

\vspace*{2cm}

\section*{Acknowledgements}
This work has been partially supported by the Spanish Ministerio de
Educaci\'on y Ciencia, MEC-FEDER project AYA2003-08739-C02-02.

\end{document}

\end{document}